\documentclass[final,5p,times,twocolumn]{elsarticle}

\usepackage{amssymb}
\usepackage{hyperref}
\usepackage[font=footnotesize,labelfont=bf]{subcaption}

\usepackage{lineno}

\journal{Nucl. Instr. Meth. A}

\begin{document}

\begin{frontmatter}



\title{Identification of light leptons and pions in the electromagnetic calorimeter of Belle II}

\author[jsi,fmful]{Anja Novosel}
\author[desy]{Abtin Narimani Charan}
\author[fmful,jsi]{Luka \v Santelj}
\author[kit]{Torben Ferber}
\author[fmful,jsi]{Peter Kri\v zan}
\author[ung,jsi]{Bo\v stjan Golob}

\address[jsi]{Jožef Stefan Institute, Ljubljana, Slovenia}

\address[fmful]{Faculty of Mathematics and Physics, University of Ljubljana, Ljubljana, Slovenia}
\address[desy]{Deutsches Elektronen-Synchrotron (DESY), Hamburg, Germany}

\address[kit]{Karlsruhe Institute of Technology (KIT) , Karlsruhe, Germany}
            
\address[ung]{University of Nova Gorica, Nova Gorica, Slovenia}
            
\begin{abstract}
The paper discusses new method for electron/pion and muon/pion separation in the Belle II detector at transverse momenta below 0.7~GeV/$c$, which is essential for efficient measurements of semi-leptonic decays of $B$ mesons with tau lepton in the final state. The method is based on the analysis of patterns in the electromagnetic calorimeter by using a Convolutional Neural Network (CNN).

\end{abstract}




\end{frontmatter}


\section{Introduction}
\label{intro}

Searches for New Physics at the intensity frontier are based on very precise measurements of rare processes within the Standard Model. Of particular interest, because of persistent hints of Lepton Flavour Universality (LFU) violation, are semi-leptonic decays of $B$ mesons, e.g. decays mediated by the $b \to c \tau^+ \nu_\tau$ transitions with a tau lepton in the final state and decays involving $b \to s \mu^+ \mu^-$ and $b \to s e^+ e^-$ transitions. In decays with tau lepton in the final state, the tau lepton must be reconstructed from its long-lived decay products, for example from the decays $\tau^- \to \mu^- \bar\nu_{\mu} \nu_{\tau}$ or $\tau^- \to e^- \bar\nu_{e} \nu_{\tau}$. In the Belle II experiment~\cite{tdr,b2-nima}, the momentum spectrum of light leptons from tau decays is rather soft, a sizable fraction being below 0.7~GeV/$c$. One of the crucial steps in the analysis of these decays is identifying low momenta light leptons ($e$ or $\mu$) from hadronic background (mostly $\pi$). The simplest baseline feature for separating electrons from other charged particles (muons and pions) is $E/p$, the ratio between the energy measured in the electromagnetic calorimeter and the reconstructed momentum of topologically matched charged track. This variable provides an excellent separation for particles with $p>1$~GeV/$c$, but due to increased energy losses from bremsstrahlung for low momentum electrons, the separation is less distinct ~\cite{b2pb}. Muons are identified in the $K_L$ and muon system. However, its efficiency is very poor for low momentum muons that are out of acceptance of the dedicated sub-detector. Other sub-detectors designed for particle identification, the time of propagation detector and the aerogel ring-imaging Cherenkov detector, are not able to provide efficient $\mu/\pi$ separation in this momentum range because at low momenta multiple scattering in the material of the detector as well as the material in front of it blurs the pattern considerably. 

Our main goal is to improve the identification of low momentum leptons using the information of energy deposition in the electromagnetic calorimeter in a form of images. As a classifier we are using a Convolutional Neural Network (CNN), a powerful machine learning technique designed for working with two-dimensional images. Using CNN on the images allows us to access the information on the shape of the energy deposition without depending on cluster reconstruction or track-cluster matching.

In what follows, we will describe the electromagnetic calorimeter of Belle II, discuss the analysis of simulated pion, muon and electron patterns in the electromagnetic calorimeter, and present the results.

\section{Electromagnetic calorimeter of Belle II}
\label{ecl}

The Belle II detector is a large-solid-angle magnetic spectrometer designed to reconstruct the products of
collisions produced by the SuperKEKB collider. The detector consists of several sub-detectors arranged
around the interaction point in cylindrical geometry: the innermost Vertex Detector (VXD) used for
reconstructing decay vertices, a combination of the Pixel Detector (PXD) and Silicon Vertex Detector
(SVD); the Central Drift Chamber (CDC) is the main tracking system; the Time of Propagation (TOP)
detector in the barrel region and the Aerogel Ring-Imaging Cherenkov detector (ARICH) in the forward endcap
region are used for hadron identification; the Electromagnetic Calorimeter (ECL) is used to measure the
energy of photons and electrons and the outermost K-Long and Muon (KLM) detector detects muons and neutral $K_L^0$ mesons \cite{tdr}.

\begin{figure}[h!]
\centering
\begin{subfigure}[h!]{\linewidth}
    \centering
    \includegraphics[width=0.49\textwidth]{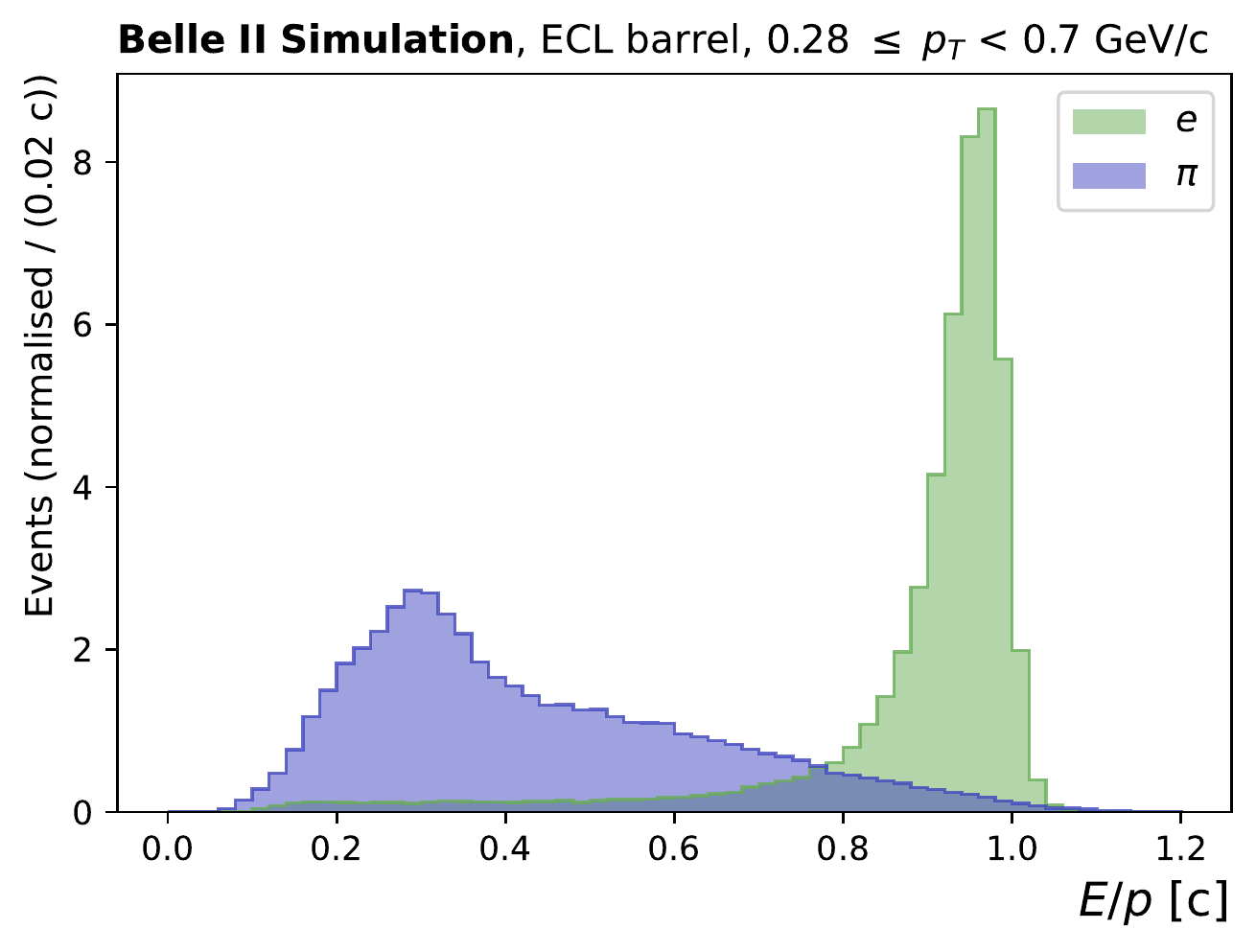}
    \includegraphics[width=0.49\textwidth]{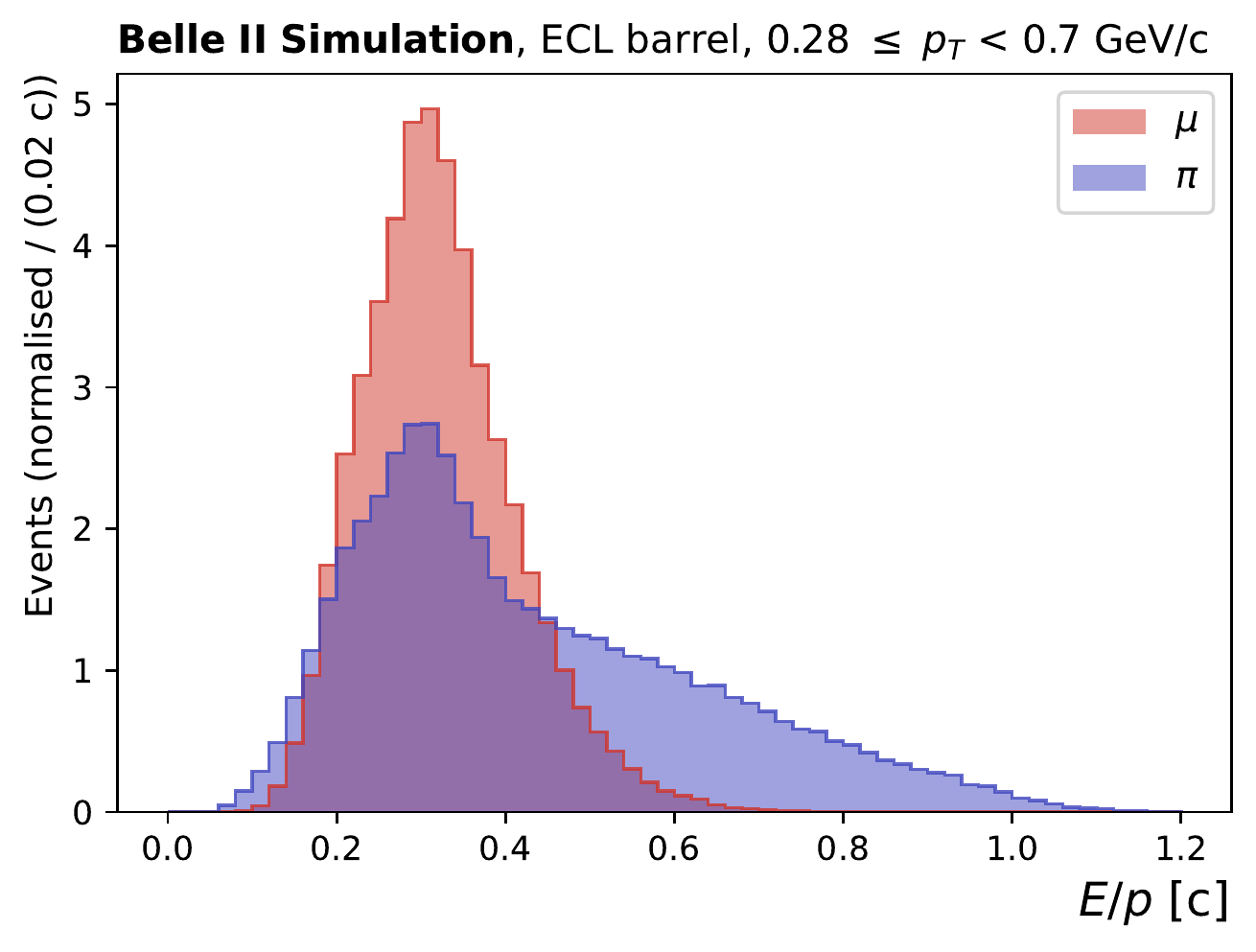}
\end{subfigure}
\caption{Distribution of $E/p$ for simulated single particle candidates: $e$ (green), $\mu$ (red) and $\pi$ (blue) for $0.28 \leq p_T < 0.7 $~GeV/$c$ in the ECL barrel region.}
\label{fig:1}
\end{figure}

The sub-detector relevant for this work is the ECL, more specifically its central barrel region barrel region which consists of 6624 CsI(Tl) scintillation crystals, covering the polar angle region $32.2 ^{\circ} < \theta < 128.7  ^{\circ}$ with respect to the beam axis. A solenoid surrounding the calorimeter generates a
uniform $1.5$~T magnetic field filling its inner volume \cite{b2-nima}. We are mainly interested in the transverse
momentum range $0.28 < p_T < 0.7$~GeV/$c$, where the minimal $p_T$ threshold ensures the tracks are within the
ECL barrel region acceptance. Currently, two methods for the particle identification in the ECL are available. The first method relies exclusively on the ratio of the energy deposited by a charged particle in the ECL and the reconstructed momentum of
topologically matched charged track, $E/p$. While for electrons this variable enables powerful discrimination, as electrons completely deposit their energy in the ECL, the $\mu/\pi$ separation is strongly limited, especially for low-momentum particles with a broader $E/p$ distribution as can be seen on Fig.~\ref{fig:1}. The second method uses Boosted Decision Trees (BDT) with several expert-engineered observables characterising the shower shape in the ECL ~\cite{lid-conf}. 

\section{Analysis of the patterns in the electromagnetic calorimeter}
\label{method}

\begin{figure}[h]
\centering
\includegraphics[width=0.48\textwidth]{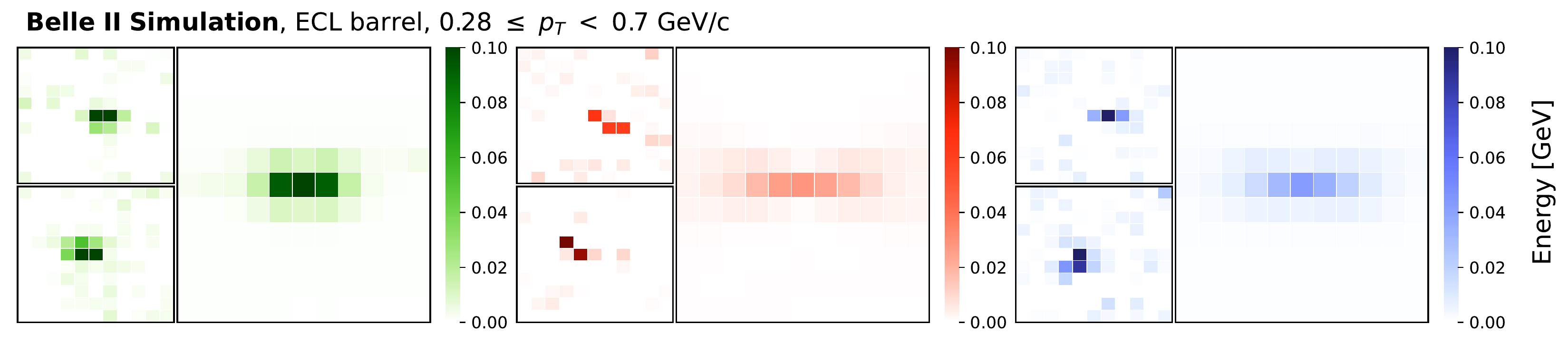}
\caption{Examples of simulated energy depositions and the average over 80000 images for $e$ (left), $\mu$ (middle) and $\pi$ (right).
}
\label{fig:2}
\end{figure}

Our proposed method to improve the identification of low-momentum leptons is to exploit the specific patterns
in the spatial distribution of energy deposition in the ECL crystals using a Convolutional Neural Network (CNN)\footnote{CNN is built using TensorFlow software available from \href{https://www.tensorflow.org}{tensorflow.org}.}. The images are consistent with the 11 x 11 neighbouring crystals around the entry point of the extrapolated track into the ECL, where each pixel corresponds to an individual ECL crystal and pixel intensity to the energy deposited by charged particle in the crystal. Examples of the obtained images are shown on Fig.~\ref{fig:2}. While electrons generate electromagnetic showers depositing the majority of their energy in the ECL, the dominant interaction in CsI(Tl) for muons and pions in the
aforementioned transverse-momentum range is ionization. Besides, pions can strongly interact with
nuclei producing less localized images compared to muons \cite{ecl-pulse-shape}.

For each binary classification we generated $1.5\times 10^6$ events using the Belle II Analysis Software Framework \cite{basf2}, where the data set consists of the same number of signal ($e$ or $\mu$) and background ($\pi$) events with uniformly distributed transverse momenta, polar angle and azimuthal angle. The two data sets were split on the train-validation-test set as $70-10-20\%$ and we use the same CNN architecture for $e/\pi$ and $\mu/\pi$ case. As an input to the convolutional layers we use 11 x 11 images. Before fully connected layers we add the information about $p_T$ and $\theta_\mathrm{ID}$, where the later represents an integer number corresponding to the location of the ECL crystal and is in the network implemented as an embedding. To perform a binary classification, we have 1 neuron in the output layer with a sigmoid activation function that outputs the signal probability that the image was produced by a lepton.

\section{Performance}
\label{perf}
\begin{figure}[h!]
\centering
\begin{subfigure}[h!]{\linewidth}
    \centering
    \includegraphics[width=0.8\textwidth]{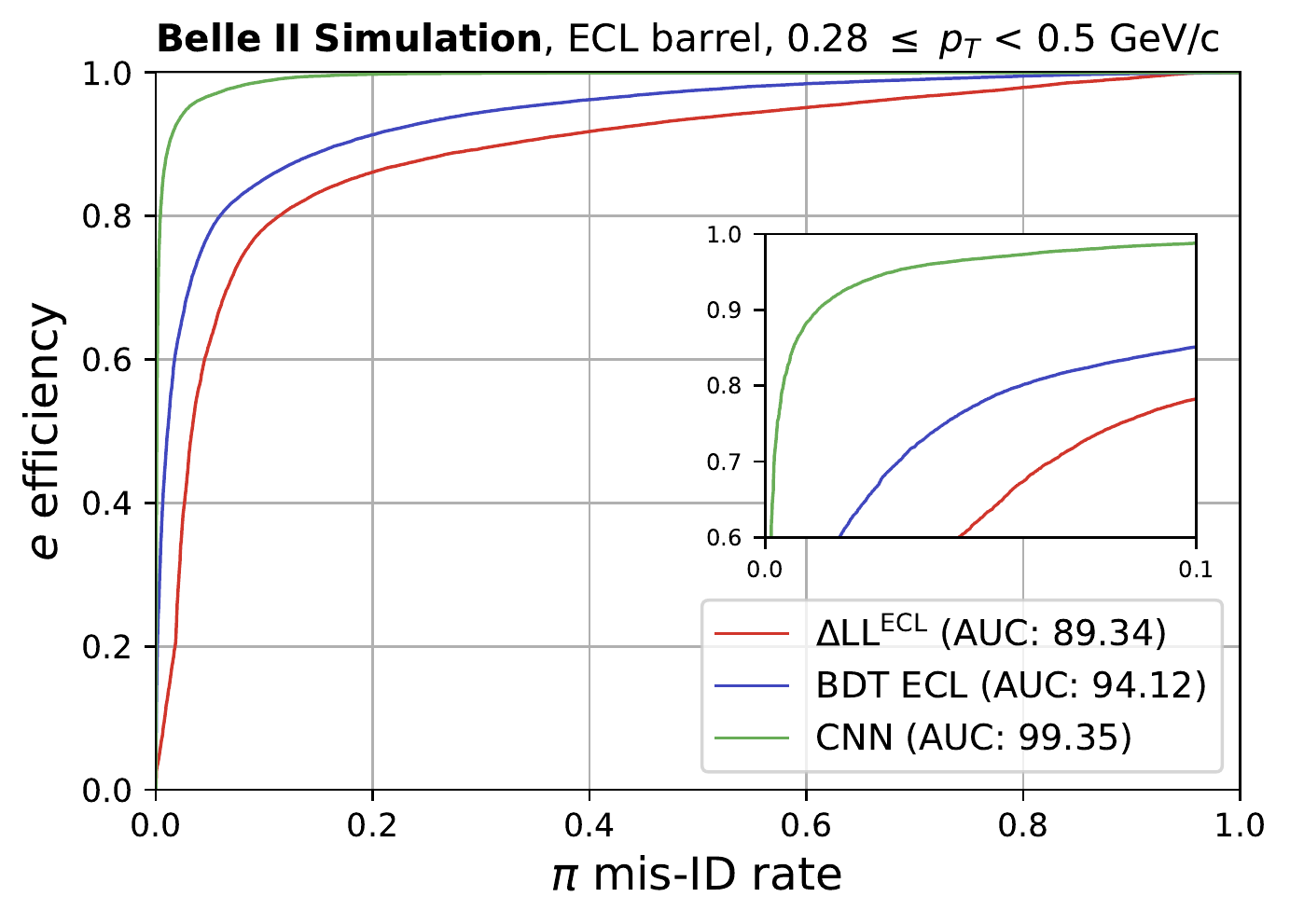}
    \includegraphics[width=0.8\textwidth]{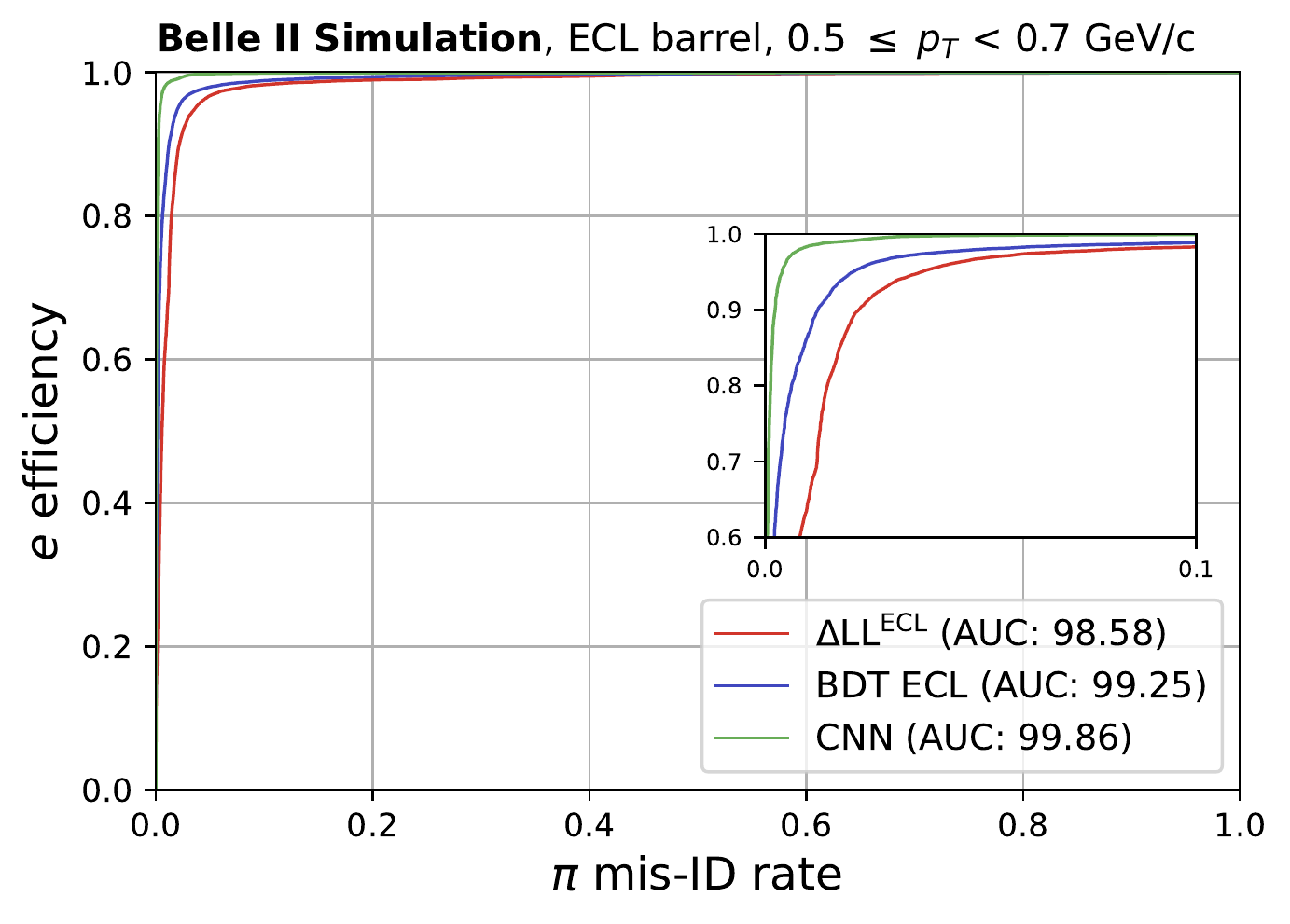}
\end{subfigure}\hfil
\caption{The performance of three different classifiers for $e/\pi$ based on only ECL information: $\Delta \mathrm{LL}^{\mathrm{ECL}}$, BDT ECL, and $\Delta \mathrm{LL}^{\mathrm{CNN}}$.}
\label{fig:3}
\end{figure}

\begin{figure}[h!]
\centering
\begin{subfigure}[h!]{\linewidth}
    \centering
    \includegraphics[width=0.8\textwidth]{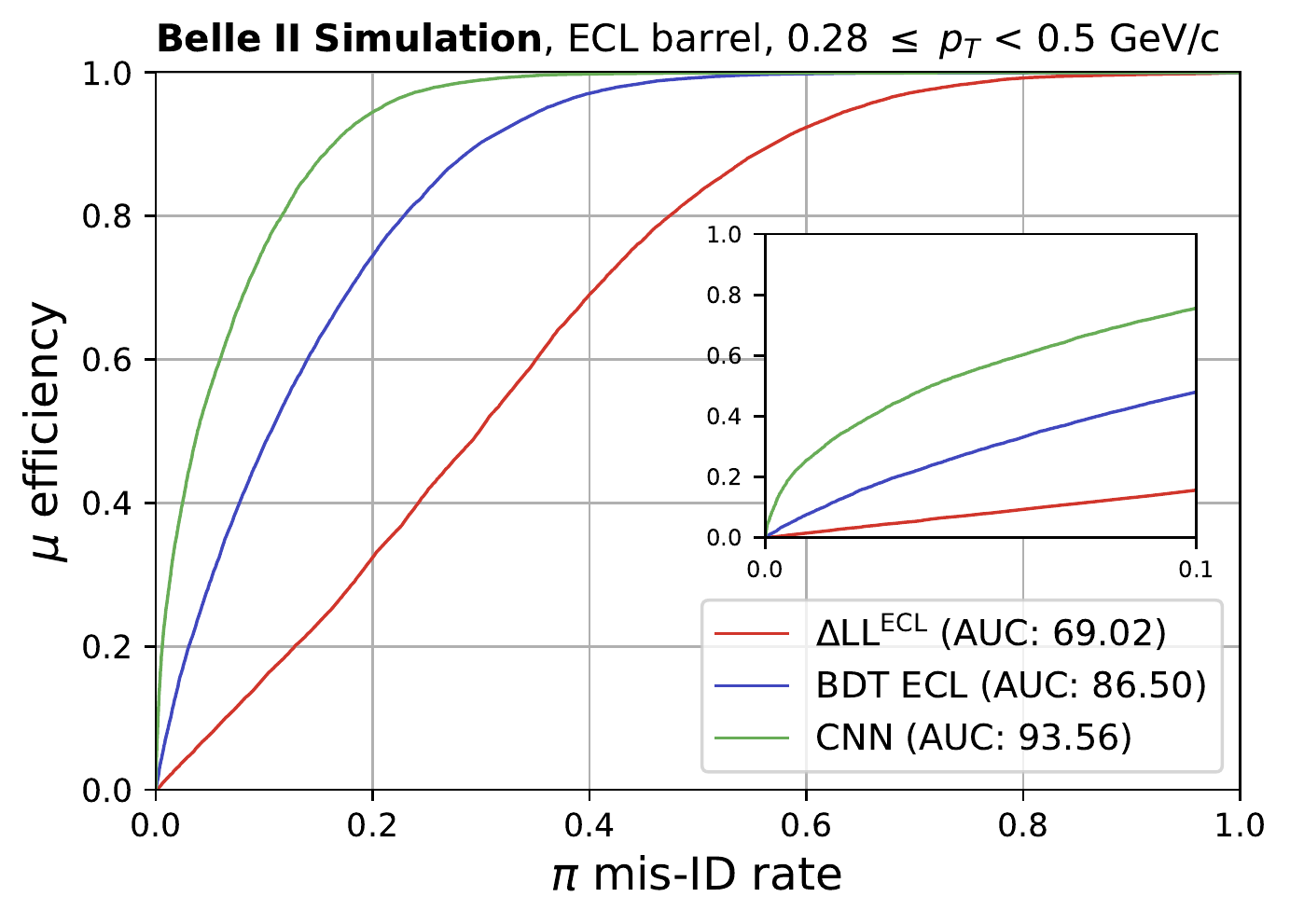}
    \includegraphics[width=0.8\textwidth]{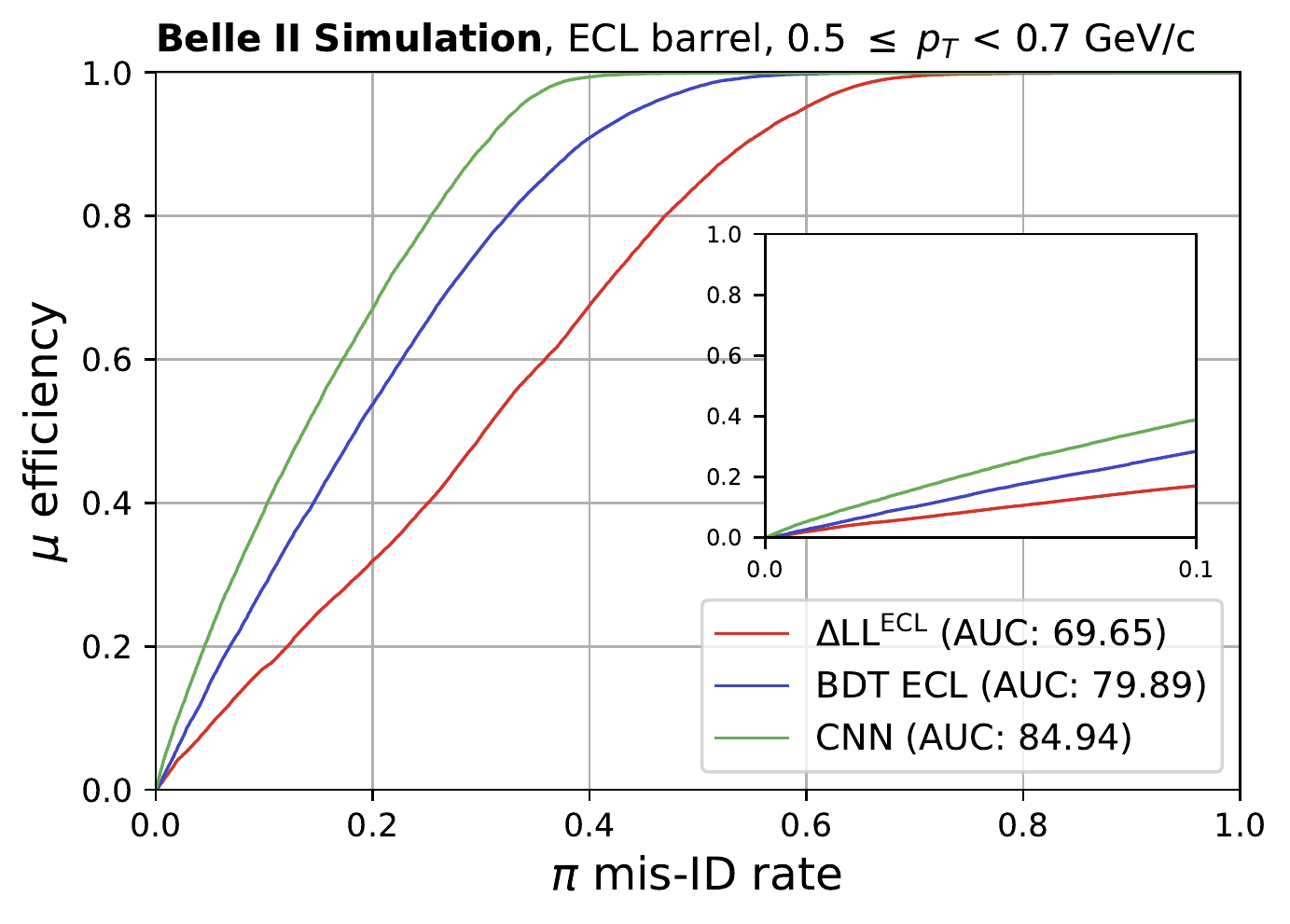}
\end{subfigure}
\caption{The performance of three different classifiers for $\mu/\pi$ based on only ECL information: $\Delta \mathrm{LL}^{\mathrm{ECL}}$, BDT ECL, and $\Delta \mathrm{LL}^{\mathrm{CNN}}$.}
\label{fig:4}
\end{figure}

To validate the performance of a binary classifier we use a Receiver Operating Characteristic (ROC) curve
by plotting true positive rate ($\mu$ or $e$ efficiency) against the false positive rate ($\pi$ mis-ID rate). As the reference for the existing ECL information, we use the log-likelihood difference, a powerful discriminator between the competing hypotheses, defined as $\Delta \mathrm{LL}^{\mathrm{ECL}} = \log{\mathrm{L}_{e,\mu}^{\mathrm{ECL}}} - \log{\mathrm{L}_{\pi}^{\mathrm{ECL}}}$ based only on $E/p$ ~\cite{b2pb} and BDT ECL using the shower-shape information from the ECL, thoroughly described in ~\cite{lid-conf}. The ROC curves obtained by these three methods are shown on Fig.~\ref{fig:3} for $e/\pi$ and on Fig.~\ref{fig:4} for $\mu/\pi$ classification.

Looking at the shapes of ROC curves and the Area Under the Curve (AUC) values, it is evident that the CNN
outperforms the existing classifiers, $\Delta \mathrm{LL}^{\mathrm{ECL}}$ and BDT ECL for both $e/\pi$ and $\mu/\pi$. The performance of the CNN drops with increasing momentum as the path in the ECL gets shorter and the specific patterns in the images become less evident. 

\section{Summary and outlook}
\label{sum}
We can conclude there is more information in the ECL that is currently used for particle identification.
We saw that the separation between low-momentum light leptons and pions can be improved using a CNN on the ECL
images. The newly proposed method looks very promising and worthwhile to be further developed. A comparison of the method presented in this article to a novel BDT-based analysis is a subject of a forthcoming publication~\cite{cnn-bdt}.   

\section{Acknowledgements}

We thank An\v ze Zupanc for his support with ideas and advice in the early stages of the project. This work was supported by the following funding sources: European Research Council, Horizon 2020
ERC-Advanced Grant No. 884719; BMBF, DFG, HGF (Germany); Slovenian Research Agency research grants No. J1-9124, J1-4358 and P1-0135 (Slovenia).

\end{document}